# Superparamagnetic iron oxide polyacrylic acid coated γ-Fe$_2$O$_3$ nanoparticles does not affect kidney function but causes acute effect on the cardiovascular function in healthy mice

Nina K. Iversen[1,2,*], Sebastian Frische[3], Karen Thomsen[2], Christoffer Laustsen[4], Michael Pedersen[4], Pernille B.L. Hansen[5], Peter Bie[5], Jérome Fresnais[6], Jean-Francois Berret[7], Erik Baatrup[1,2] and Tobias Wang[1]

[1]Zoophysiology, Department of Biological Sciences, Aarhus University
[2]Interdisciplinary Nanoscience Center, Aarhus University
[3]Department of Biomedicine, Aarhus University
[4]MR Research Center, Aarhus University Hospital, Aarhus University
[5]Cardiovascular and Renal Research, Institute of Medical Biology, University of Southern Denmark,
[6]Physicochimie des Electrolytes, Colloïdes et Sciences Analytiques (PECSA) UMR 7195 CNRS-UPMC-ESPCI, 4 place Jussieu, 75252 Paris Cedex 05, France
[7]Matière et Systèmes Complexes, UMR 7057 CNRS Université Denis Diderot Paris-VII, Bâtiment Condorcet, 10 rue Alice Domon et Léonie Duquet, 75205 Paris, France

Corresponding author
*Nina K. Iversen
Department of Bioscience
C.F. Moellers alle 3, building 1131
Aarhus University
8000 Aarhus C
Denmark

**Abstract**
This study describes the distribution of intravenously injected polyacrylic acid (PAA) coated γ-Fe$_2$O$_3$ NPs (10 mg kg$^{-1}$) at the organ, cellular and subcellular levels in healthy BALB/cJ mice and in parallel addresses the effects of NP injection on kidney function, blood pressure and vascular contractility. Magnetic resonance imaging (MRI) and transmission electron microscopy (TEM) showed accumulation of NPs in the liver within 1h after intravenous infusion, accommodated by intracellular uptake in endothelial and Kupffer cells with subsequent intracellular uptake in renal cells, particularly the cytoplasm of the proximal tubule, in podocytes and mesangial cells. The renofunctional effects of NPs were evaluated by arterial acid-base status and measurements of glomerular filtration rate (GFR) after instrumentation with chronically indwelling catheters. Arterial pH was 7.46±0.02 and 7.41±0.02 in mice 0.5h after injections of saline or NP, and did not change over the next 12h. In addition, the injections of NP did not affect arterial PCO$_2$ or [HCO$_3^-$] either. Twenty-four and 96h after NP injections, the GFR averaged 11.0±0.8 and 13.0±0.4 ml min$^{-1}$ g$^{-1}$, respectively, values which were statistically comparable with controls (14.0±0.2 and 14.0±0.6 ml min$^{-1}$ g$^{-1}$). Mean arterial blood pressure (MAP) decreased 12-24h after NP injections (111.1±11.5 vs 123.0±6.1 min$^{-1}$) associated with a decreased contractility of small mesenteric arteries revealed by myography to characterise endothelial function. In conclusion, our study demonstrates that accumulation of superparamagnetic iron oxide nanoparticles does not affect kidney function in healthy mice but temporarily decreases blood pressure.

# Introduction

In recent years, the vast clinical potential of nanomedicine has incited the development of various multimodality particles with distinct biological, physical and chemical properties useful for imaging, as well as various biomedical and therapeutic applications (Farokhad and Langer, 2006; Gu, 2007; Longmire et al., 2008; Liong et al., 2008). Superparamagnetic iron oxide nanoparticles (NP) have been explored for targeted drug delivery, hyperthermia, tissue repair, cell sorting, and contrast agents for magnetic resonance imaging (Rosen et al., 2012; Jain, 2007; Gupta et al., 2007). The NPs are often chemically modified and synthesised to increase stability and safety. In contrast to conventional imaging agents, NPs are relatively stable *in vivo*, exemplified by the recent findings that quantum dots are retained in the body for 100 days (Choi et al., 2007) and even up to two years (Ballou et al. 2007). It is imperative, therefore, to study the specific distribution of magnetic NP to understand the specificity and long-term biodistribution profiles to access a better clinical efficiency and safety (Rosen et al., 2012). It has been shown that excessive release of free ions from NPs that may be toxic (Weir et al. 1984), lead to oxidative stress (Hussain et al., 2005; McCord 1996) and disturb liver metabolism (Wisse et al. 1991). In addition, cirrhosis and hepatocellular carcinoma can develop when the liver iron concentration exceeds 4 µg $g^{-1}$ wet weight (Neuberger et al. 2005). Hussain et al. (2005) recently demonstrated that high concentrations of $F_3O_4$ NP *in vitro* reduced cell proliferation and caused cell death in rat liver cells.

Inhalation of air-borne particular matter (PM) increases mortality and morbidity from pulmonary mediated cardiac arrhythmias (Samet et al. 1994; Zhu et al. 2008) that seems to be precipitated by disrupted autonomic cardiovascular regulation (Magari et al. 2001). PMs are a complex mixture of organic and inorganic chemicals, including metals and particulates (Mo et al. 2009) and particularly the ultra–fine particles (UFPs) with an aerodynamic diameter less than 100 µm seem to elicit cardiorespiratory malfunctions (Stern and McNeil 2008). Recent human and animal studies show that inhaled UFPs swiftly enter the systemic circulation (Nemmar et al. 2001; Pope III et al. 2004), where they disturb myocardial or vascular endothelial cell function (Pope III et al. 2004; Dockery et al. 1993; Pekkanen et al. 2002) by generation of reactive oxygen species (ROS; Mo et al. 2009). Nevertheless, the specific biological interactions leading to the possible cardiovascular effects of PM and UFP remain largely unknown (Pekkanen et al. 2002) precluding successful mitigation of their negative impact. Many of the NPs explored for biomedical use share some physicochemical properties with UFPs and are therefor also believed to have similar detrimental effects on the cardiovascular system (Oberdörster et al. 2002).

Considerable research has been devoted to understand and characterize the distribution and accumulation of various metal and magnetic nanoparticles in healthy organs, but less attention has been drawn to elucidate renal effects following NP injection. It is known that engineered particles, such as iodinated contrast agents, can cause acute kidney injury (Byrd and Sherman 1979; Nash et al. 2002; Bruce et al. 2009; Cochran et al., 1983) and the incidence of acute kidney injury caused by contrast administration is commonly referred to as contrast-induced nephrotoxicity (Nash et al., 2002). Intrarenal accumulation of NPs may lead to acute kidney injury (Byrd and Sherman 1979; Cochran et al. 1983; Bruce et al. 2009) followed by a more permanent renal failure (Toprak 2007), likely mediated by medullary hypoxia due to decreased renal blood flow, secondary to vasoconstriction, tubular obstruction, direct tubular toxicity and oxidative damage (Persson and Tepel 2006). Furthermore, Chen et al. (2006) observed that mice exposed to nano-copper developed glomerulitis, degeneration and necrosis of renal tubules, and renal inflammation.

The objectives of this study were to determine: a) the destiny of intravenously injected polyacrylic acid (PAA) coated $\gamma$-$Fe_2O_3$ particles at a clinical relevant dose (Ma et al., 2012) for hyperthermic cancer therapy with particular emphasis on the renal and hepatic accumulation, b) the renal function (systemic acid/base status and GFR) following exposure to these particular NPs and c) the effect of the NPs on cardiovascular functions (MAP and vascular contractility).

# Methods

*Experimental animals*

BALB/cJ mice of both sexes (7-10 weeks) were purchased from Taconic or Harlan (Denmark) and transported to the animal facility at Department of Biological Sciences, Aarhus University. In total 104 mice were used divided into the following studies to address the a) distribution of polyacrylic acid (PAA) coated $\gamma$-$Fe_2O_3$ particles (MR n=9, TEM n=8), b) the renal function (systemic acid/base status n= 21 and GFR n=18), c) cardiovascular effects (MAP n=12 and vascular contractility, n=36).
The mice used for GFR measurements were housed at the Biomedical Laboratory, University of Southern Denmark. At both institutions, mice were kept under standard light (12:12 dark-light) and temperature conditions with free access to rodent chow and tap water. Animal care followed the guidelines of the National Institutes of Health and the experimental protocol was approved by the Danish Animal Experiments Inspectorate.

*Magnetic resonance imaging*

A clinical 1.5 T Phillips Achiva MRI system (Philips Medical Systems, Best, Netherlands) was used to visualize the *in vivo* distribution of NPs. NPs were administered directly to the tail vein 1 h (n=3) or 96 h (n=6) prior to MRI, while a third group of three mice received a sham injection of saline. All mice were euthanized with an over-dose of pentobarbital (10 mg $kg^{-1}$) immediately before positioned in the scanner.

The iron-containing NPs cause a dephasing of the nearby spins of water, thereby decreasing the signal intensity on a T2*-weighted images. Thus, we applied a standard T2*-weighted gradient-echo sequence using the following parameters: field-of-view = 260x260 $mm^2$, matrix = 384x384, 10 slices with a thickness of 1 mm, repetition time = 800 ms, echo time = 20 ms, excitation angle = 90, and number of averages nt = 3. A spin-echo sequence, for T2 relaxation mapping, was applied: field-of-view = 260x260 $mm^2$, matrix = 384x384, 10 slices with a thickness of 1 mm, repetition time = 2000 ms, array of echo times = 20,40, 60, 80, 100, 120, 140, 160 ms, and number of averages nt = 2.

All data were exported in DICOM format to the Mistar (Apollo Imaging Technology, Melbourne, Australia) and ImageJ (a public domain, Java-based image processing program developed at the National Institutes of Health) analysis software. Relaxometric analysis was used to generate parametric T2 map based on the acquired multi-echo spin-echo images. A global intensity profile was computed for all 10 slices, and NP accumulation was confirmed by reduction in signal intensity in kidney and liver regions.

*Transmission electron microscopy (TEM).*

$\gamma$-$Fe_2O_3$ particles (10 mg $kg^{-1}$) were given in the venous catheter in two experimental groups with different exposure duration (24h and 96h, n= 3 in each), while additionally two mice received saline and thereby served as controls. After 24h and 96h, respectively, the mice were deeply anaesthetised by intraperitoneal injections of pentobarbital (50 mg ml$^{-1}$, 0.01 ml g$^{-1}$) and perfused transcardially with a 2 % paraformaldehyde and 2.5 % glutaraldehyde fixative in a 0.1 M phosphate buffer (pH 7.4) for 5 min until all organs became pale. Liver and kidney were transferred to individual glass vials with the fixative and subsequently post fixated in 4% osmium tetroxide dissolved in milli Q water. After 24 h, the tissue blocks were dehydrated in a series of increasing concentrations of ethanol (70, 90, 96 and 99.9 %), immersed in the organic intermedium propylene oxide and embedded in EMBed-812 (Electron Microscopy Sciences, Hatfiled, PA). Ultrathin sections for TEM were cut on a Leica Ultracut UCT ultramicrotome (Leica Mikrosysteme GmbH, Vienna Austria) with a 45 degrees angled diamond knife (Diatome, Biel, Switzerland). The ultrathin sections were stained with uranylacetate and inspected in a CM 100 FEI transmission electron microscope (FEI, North America NanoPort, Hillsboro, Oregon 97124, USA) and photographed with a CCD camera (1K MegaView, Olympus Soft Imaging Solutions GmbH, Münster, Germany), followed by image processing in AnalySIS (Olympus, Germany).

Electron Energy Loss Spectroscopy (EELS; Gatan Quantum SE/963 spectroscope and camera (Gatan, Inc. Pieasanton, CA 94588) on a Titan Krios 80-300 microscope ( FEI, North America NanoPort, Hillsboro, Oregon 97124 USA)) were used to determine the elemental constitution of the particles imaged in TEM. The presence of the $Fe_2O_3$ in the particles was evaluated by comparing the position of the Fe-L2 and Fe-L3 edges in the obtained EELS spectrum with the positions of Fe-L2 and Fe-L3 edges from $Fe_2O_3$ in the EELS Atlas. A 30 eV slit was applied for acquisition of pre-edge (slit center 683 eV) and post-edge (slit center 728 eV) images. Jump-ratio was calculated from these recordings using Gatan software. The size of the particles was measured using ImageJ on the TEM-images from the for EELS analysis.

*Insertion of arterial and venous catheters for GFR measurements*

The eighteen mice used for determination of glomerular filtration rate (GFR) measured by inulin clearance, were anaesthetised by an intraperitoneal injection of a combination drug (Ketalar/Rompun or Ketamin/Xylazine). Chronic indwelling catheters were placed in the femoral artery and vein for measurements of arterial blood pressure and venous infusions, respectively (Mattson, 1998). Micro-Renathane Tubing MRE-040 and MRE-033 (AgnTho's AB; Sweden) were used as catheter tips. The catheters were exteriorized through a subcutaneous tunnel from the groin to the neck where additionally a lightweight tethering spring to a steel button in the neck was implanted. Following the operation, the spring containing the catheters was secured to a swivel above the cage enabling the mice to move unrestrained. To maintain catheter patency, a continuous infusion of 10 µl h$^{-1}$ of heparin diluted isotonic glucose to 100 IU ml$^{-1}$ was started in the arterial side. Twice within the first 24h after surgery, the mice were given subcutaneous injections of Temgesic for analgesia (Buprenorphinum, 0.3 mg ml$^{-1}$, 3.75 mg kg$^{-1}$). The mice were allowed to recover for four days before the experiments were conducted.

The actual experiments were initiated by connecting the arterial line to a pressure transducer (Föhr Medical Instruments, Hessen, Germany), and data were collected at 200 Hz using Lab View software (National Instruments, Austin, TX, USA).

*Glomerular filtration rate (GFR)*

Four days after surgery, the mice had fully recovered and no change in blood pressure or heart rate was observed the following days. Basal blood pressure was measured for 1h before a consecutive 100µl bolus infusion of NP in nine mice through the venous catheter. As control, 100µl bolus infusion of saline was given to another nine mice. Initially, a 50µl isotonic bolus was given to place the saline in the dead space of the catheter. Inulin (Inutest 25%) was infused at the venous side at 10 µl h$^{-1}$ for 24 and 96 hours to achieve steady state. Then a small blood sample (100 µl) was withdrawn from the arterial catheter at both time points.

Plasma Inulin concentration was determined by Atomic Absorption Spectrometry. The plasma samples were diluted six times in 10 mM Na-Pi suspension (pH 5.0) and incubated with 50 µl inulinase pr 50 µl sample for at 37 °C for 10 min. Then 130 µl SDH/NADH was added and the suspension was further incubated for 60 min. The absorbance was determined at the wavelengths at 340-490 nm. GFR was calculated as inulin infusion rate / plasma inulin concentration.

*Insertion of arterial for blood sampling*

To sample arterial blood from undisturbed mice, catheters were inserted into the left carotid artery. Catheters were also inserted in the left jugular vein for injection of NPs in some experiments. During surgery, the mice were anaesthetised with an intraperitoneal injection of ketamine and xylazine (75 and 10 mg kg$^{-1}$, respectively; 0.01 ml g$^{-1}$ mouse). A small thermometer was inserted into the rectum, so that body temperature could be maintained at 36-37.5 °C throughout surgery.

Under a dissecting microscope (Olympus, SZ-STB1, Japan), the left carotid artery or the left jugular vein were cannulated with stretched PE-10 tubing (Portex fine bore polythene tubing, Smiths Medical ASP, Inc., Keene, USA) containing heparinized saline (50 IU/ml, 0.9 % NaCl). The catheter was led through a small incision in the back of the neck, exteriorized behind the ear and secured to the skin by a single stitch. All mice received a subcutaneous injection of 1 ml saline to prevent dehydration. The mice were allowed to recover for at least 24 h at 32°C after surgery and received two subcutaneous injections of Temgesic for analgesia (Buprenorphinum, 0.3 mg ml$^{-1}$, 3.75 mg kg$^{-1}$).

*Arterial blood gases*

Acid-base parameters were measured 30 min, 3h and 12h after NP had been administrated through the tail vein (n=8 in all groups). In each group, four mice received NPs while the other four received saline injected through the tail vein. To measure blood gases, a 300 µl of arterial blood was taken through the catheter of undisturbed mice.

Blood pH was measured with a Radiometer (Copenhagen, Denmark) capillary pH electrode thermostated to 37°C and connected to a Radiometer PHM 73. Total plasma $CO_2$ concentration ($[CO_2]pl$) was measured as described by Cameron (Cameron, 1971), and plasma bicarbonate concentration ($[HCO_3^-]pl$) was calculated as $[CO_2]pl - (\alpha CO_2 \times PCO_2)$, using a plasma $CO_2$ solubility ($\alpha CO_2$) at 37°C of 0.0301 mmol l$^{-1}$ mmHg$^{-1}$ (Siggaard-Andersen (1976).

Plasma $[HCO_3^-]$ was calculated from pH and total plasma $[CO_2]pl$ using the Henderson–Hasselbalch equation. The $CO_2$ solubility coefficient of 0.0301 mmol l$^{-1}$ mmHg$^{-1}$ was taken from Siggaard-Andersen (1976) and pK′ was considered to be 7.816 ±0.234pH (Iversen et al., 2012). Haematocrit was measured in duplicate after centrifuging blood in micro-haematocrit tubes at 12·000·r.p.m. for 3·min.

*Surgical implants of telemetric probes*

Surgical anaesthesia was induced by an intraperitoneal injection of ketamine (75 mg kg$^{-1}$) and xylazine (10 mg kg$^{-1}$). A small thermistor was inserted into the rectum to maintain body temperature at 36-37.5 °C throughout surgery, and the eyes were protected against dehydration with an eye ointment (Ophtha A/S, Copenhagen, Denmark). After the neck had been shaved and disinfected, the mouse was placed in a supine position and a ventral midline skin incision was made from the lower mandible posteriorly to the sternum (3 cm). The sub-maxillary glands were gently separated and the left common carotid artery was isolated using fine forceps, without disrupting the vagus nerve parallel to the artery. Two sterile silk sutures (6-0, Agnthos, Sweden) were passed under the vessel for temporary occlusion of the artery during catherisation and to secure the catheter. The catheter from the telemetric probe (TA11PA-C10: 1.1 cm 3, 1.4 g; Data Sciences International (DSI), St Paul, MN, USA) was inserted into the artery, and a subcutaneous pouch was formed for placement of the transmitter body along the animal's right flank. The neck incision was closed using 5-0 silk. Mice were allowed to recover from surgery in a climate chamber at 28 ºC the first 24h after surgery. Twice within this re-covalence period, the mice received subcutaneous injections of Temgesic for analgesia (Buprenorphinum, 0.3 mg ml$^{-1}$, 3.75 mg kg$^{-1}$).

*Telemetry protocol*

Following recovery from surgery and anesthesia, the 12 mice instrumented with the telemetric blood pressure probes were returned to their cages (placed on top of the telemetry receivers), and monitored daily throughout the study. Radio telemeters (Data Sciences International (DSI), St Paul, MN, USA) were magnetically activated 7 days after surgery, and blood pressure (pulsatile waveforms and mean arterial pressure, (MAP)), heart rate, and locomotor activity were recorded continuously as 10 sec averages each min for 7 days. After recording cardiovascular parameters for 24h, NP ($\gamma$-Fe$_2$O$_3$) or saline were injected in the tail vein. The telemetered pressure signals (absolute pressure) were corrected automatically for changes in atmospheric pressure measured by an ambient pressure monitor.

*Isometric force measurements in mesenteric small arteries*

Thirty-six mice of both sexes were infused with either NPs or saline in the tail vein and mesenteric arteries were removed at different time points for *in vitro* studies. Mice that received $\gamma$-Fe$_2$O$_3$ (100 µl; 10 mg kg$^{-1}$) were killed after 24 h, 72h and 7 days (n=6 in each groups), respectively, while control animals having received saline were killed at the same points in time as the treated mice. Twelve of the arteries harvested from the control group were exposed to $\gamma$-Fe$_2$O$_3$ NPs (10 or 50 mg kg$^{-1}$) for 30 min *in vitro*.

In all mice, the mesenteric bed was carefully pinned out in a Petric dish and four 2 mm long segments of third-order branches were dissected free from fat and connective tissue (Mulvany et al., 1978). The isolated arteries were threaded onto two stainless steel wires (40 µm in diameter) and mounted as ring preparations in a quadruple wire myograph (Model 610M, Danish Myo Technology, Aarhus University) for isometric force measurements. Data were sampled at 100 Hz (16 bit) using a Biopack MP100 data acquisitions system (Biopac system Inc. Santababara, CA, USA). Upon mounting, the arteries were allowed to equilibrate for 20 min before they were stretched stepwise to characterize the passive elastic properties (Mulvany and Halpern, 1977). The mesenteric arteries were stretched to 90 % of L$_{100}$, where L$_{100}$ is defined as the circumference of the relaxed artery at a transmural pressure of 100 mmHg. Prior to the experimental protocol, the arteries were stimulated three times with 10µM noradrenaline (NA) dissolved in 125 mM K$^+$-containing physiological saltwater solution (PSSS (KPSS)) for 3 min to verify viability. Subsequently, NA (3x10$^{-3}$) was added and at the maximal response, a single dose of acetylcholine (ACh; 1 µM) was given to ensure relaxation of the vessel. This was repeated with an increased dose of Ach to 3 µM. Vessels that failed to contract more than 125 % of the forced developed at L$_{100}$ were discarded. Both NA and ACh were added directly to the bath. To evaluate the contractile properties, we constructed cumulative NA concentration-response curves (0.1-0.3 µM NA), increasing the concentration every 2 min to allow the force to reach a steady state.

The vessels were dissected, mounted and, if not indicated otherwise, held relaxed in physiological salt solution (PSS) of the following composition in mM: NaCl, 119; NaHCO$_3$, 25; KCl 4.9; CaCl$_2$, 2.5; H$_2$PO$_4$, 1.18; MgSO$_4$, 1.17; EDTA, 0.026, and glucose 11. KPSS was PSS in which Na$^+$ was substituted by an equimolar concentration of K$^+$ to reach 15 mM K$^+$. All solutions were bubbled with 5% $CO_2$ in $O_2$ at 37°C and pH was adjusted to 7.4 immediately before use.

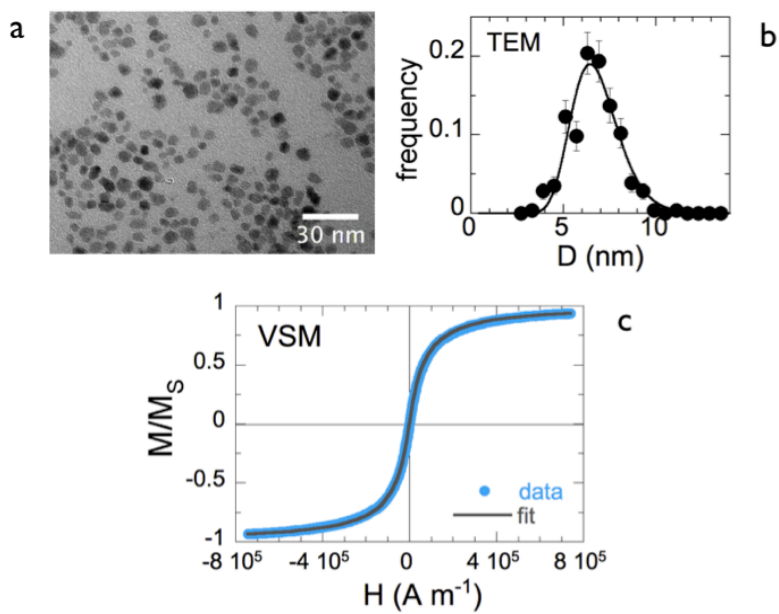

**Figure 1** *Characteristics of the γ-Fe$_2$O$_3$ NP synthesized for this study. A: Transmission electron microscopy (TEM) image of uncoated γ-Fe$_2$O$_3$ NP particles, B: Size distribution of γ-Fe$_2$O$_3$ NP nanoparticles obtained by TEM. The continuous line is a log-normal function with median diameter 6.8 nm and polydispersity 0.18, C: Magnetic field dependence of the macroscopic magnetization normalized by its saturation value for a 10 wt.% g-Fe$_2$O$_3$ dispersion. The solid curve results from the convolution between the Langevin equation for paramagnetism and a log-normal function with median diameter 6.7 nm and polydispersity 0.21.*

*Nanoparticles*

Nano-sized γ-Fe$_2$O$_3$ coated with poly(acrylic acid) 2000 g mol$^{-1}$ (PAA$_{2K}$) (Laboratoire Matière et Systèmes Complexes, Université Denis Diderot Paris-VII) were used in this experiment (Fig. 1). The synthesis of iron oxide nanoparticles (γ-Fe$_2$O$_3$, maghemite) was based on the polycondensation of metallic salts in alkaline aqueous media (Massart et al., 1995). A transmission electron microscopy (TEM) image obtained from a dilute aqueous dispersion exhibited compact and spherical particles (Fig. 1 A). The size distribution of these particles was described by a log-normal function with a median diameter $D_0^{TEM} = 6.8\,nm$ and polydispersity of $s^{TEM} = 0.18$ (Fig. 1 B). The polydispersity is described as the ratio between standard deviation and the mean value. Vibrating sample magnetometry (VSM) was carried out to determine the magnetization *versus* excitation from a dispersion at concentration 10 wt. %. Fig. 1 C shows the evolution of the macroscopic magnetization $M(H)$ normalized by its saturation value $M_S$. The solid line in grey was calculated using the Langevin function of paramagnetism convoluted with a log-normal distribution function of the particle sizes. The parameters of this distribution were $D_0^{VSM} = 6.7\,nm$ and $s^{VSM} = 0.21$, in good agreement with the those obtained by TEM. Further characterization of the particles including the determination of the surface charge and structural anisotropy can be found in our previous work (Berret et al., 2007). To improve their colloidal stability, the cationic particles were coated with M$_W$ = 2000 g mol$^{-1}$ poly(acrylic acid) using the precipitation-redispersion process. This process resulted in the adsorption of a highly resilient 3 nm polymer layer surrounding the particles. The hydrodynamic diameters of the bare and coated NPs were 12.7 and 17.7 nm, respectively (Berret et al., 2007; Chanteau et al., 2009; Safi et al., 2010; 2011).

*Statistical analysis*

Data were tested for homogeneity of variance and normal distribution prior to parametric tests. Statistically significance of differences between treatments on blood pressure and heart rate obtained in mice receiving PAA coated γ-$Fe_2O_3$ NP or saline were evaluated by a two-factor (time and treatment) analysis of variance (ANOVA) for repeated-measures followed by a Tukey multiple-comparison (Sigmaplot and Statistic 11.0). Effects of treatment with NP and the time period exposure on GFR were tested by a t-test. Differences between means were considered statistically significant when $P≤0.05$ and all results are presented as means ± S.E.M.

The telemetry data were collected continuously by sampling for 10 sec every minute and stored using the Dataquest ART data acquisition system (Data Sciences International). After the experimental session, the obtained data was uploaded into the ACQ analysis program (Data Sciences International) and exported into Excel. The data from the myograph was collected at 50 Hz and transferred directly into Excel. All further data analyses from both experimental sessions were analyses in Excel. Data were tested for homogeneity of variance and normal distribution prior to parametric tests. Statistical significance of differences between treatments on blood pressure, heart rate or locomotor activity obtained in mice receiving PAA coated γ-$Fe_2O_3$ NP or saline was evaluated by a two-factor (time and treatment) analysis of variance (ANOVA) for repeated-measures followed by a Tukey multiple-comparison. In the myograph studies, the effect of treatment, i.e. PAA coated γ-$Fe_2O_3$ NP or saline and the cumulative dose-response curve of noradrenalin on tension and pressure was also evaluated by a two-factor (noradrenalin and treatment) analysis of variance (ANOVA) for repeated-measures followed by a Tukey multiple-comparison. Differences between means were considered statistically significant when $P≤0.05$ and all results are presented as means ± S.E.M.

# Results

The overall distribution of the intravenously injected PAA coated γ-$Fe_2O_3$ NPs (Fig. 1) in healthy mice was revealed by MRI (Fig. 2). Because these particles are magnetic, they disturb the local magnetic field in the tissue. Therefore, the accumulation of NPs is indirectly recognized as decreased signal intensity in the MRI images. The accumulation of PAA coated γ-$Fe_2O_3$ NPs was visible in the liver within 1h after intravenous infusion (Fig 2 B) and subsequent accumulation in kidney and spleen 3h after injections (Fig 2 C, F). This distribution pattern did not change over the next 96h (Fig. 2 C, F). Transmission Electron Microscopy (TEM) confirmed the presence of these NPs in the liver and kidney. The particles were very clearly present in vesicles, most likely belonging to the endocytic/lysosomal system, in the cytoplasm of endothelial (Fig 3 A and B) and Kuppfer cells in the liver sinusoids (Fig. 3 C and D). In the kidney, the TEM images showed the NPs to be present in vesicles in the cytoplasm most likely belonging to the endocytic/lysosomal pathway in podocytes (Fig. 4 A, B, C and D), proximal tubule cells (Fig. 4 E and F), and mesangial cells (not shown). Electron Energy Loss Spectroscopy (EELS) showed edges at positions consistent with Fe-L2 and Fe-L3 edges from $Fe_2O_3$ (Fig. 5 A and B). The diameter of the particles was 6.24± 0.18 nm (n = 30). Jump-ratio analysis confirmed the Fe specific edges to origin from the particles (Fig. 5 C, D and E). These results confirms that the particles imaged by TEM consist of $Fe_2O_3$, and thus to be the injected NPs.

*Arterial acid-base status*

To assess whether the NP deposition in the kidneys (Fig. 2, 4), particularly proximal tubular cells, would affect or disturb normal arterial acid-base status, we measured arterial pH, PCO$_2$ and [HCO$_3^-$] at 0.5h, 3h and 12h after intravenous injections of γ-Fe$_2$O$_3$NP or saline (Table 1). Arterial pH was statistically similar 0.5h after injections of saline or NP in mice (7.46±0.02 and 7.41±0.02 respectively). These values remained unchanged over the next 12h. Injections of NP did not affect PCO$_2$. There was a slight (but statistically insignificant) tendency of decrease in [HCO$_3^-$] 3h after injections of NP (Table 1), but the concentration after 12h was similar to the 0.5h experimental group. In conclusion, neither pH, PCO$_2$ nor [HCO$_3^-$] were affected of saline injections or significantly different from the experimental groups receiving NP.

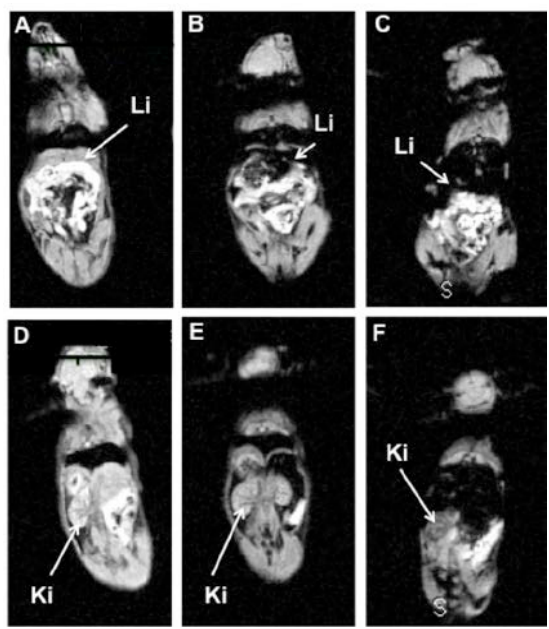

*Figure 2* *Magnetic resonance imaging of the distribution of intravenously injected PAA coated γ-Fe$_2$O$_3$ NP (10 mg kg$^{-1}$) A: representative control mouse, B: 1h after NP administration, liver, C: 96h after administration, liver, D: representative control mouse, E: 1h after NP administration, kidney, C: 96h after administration, kidney. Since the present NPs are magnetic, they will disrupt the magnetic signals in the MR scanner, where as NP build up are seen as dark areas.*

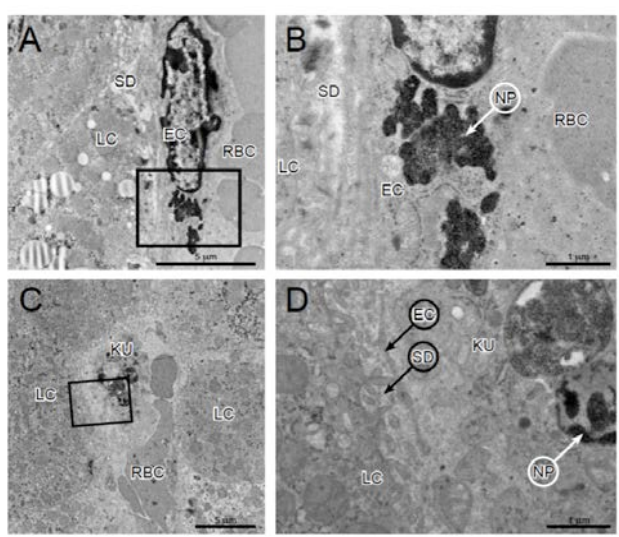

*Figure 3* *Transmission electron microscopy (TEM) of PAA coated γ-Fe$_2$O$_3$ NP accumulated in the kidney. The NPs are seen as small dark spots. BB: Brushborder, L: Lysosome, EC: endothelial cell, M: Mitochondria, RBC: Red blood cell, PC: Podocyte, GBM: glomerular basal membrane, NP: Nanoparticles*

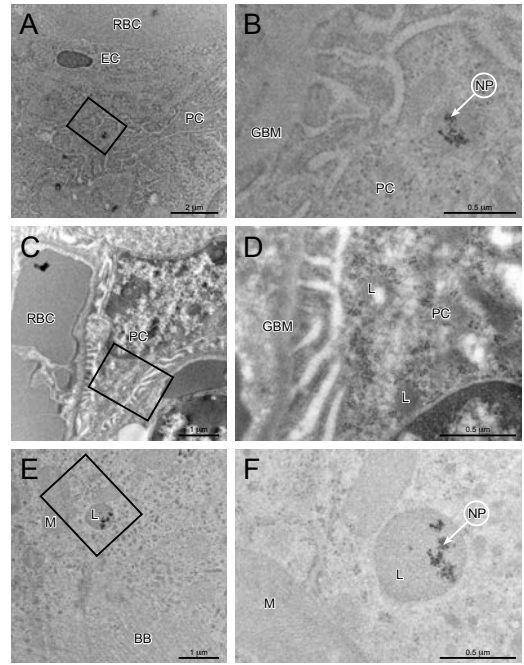

*Figure 4* Transmission electron microscopy (TEM) of PAA coated γ-Fe$_2$O$_3$ NP accumulated in the kidney. The NPs are seen as small dark spots. BB: Brushborder, L: Lysosome, EC: endothelial cell, M: Mitochondria, RBC: Red blood cell, PC: Podocyte, GBM: glomerular basal membrane, NP: Nanoparticles

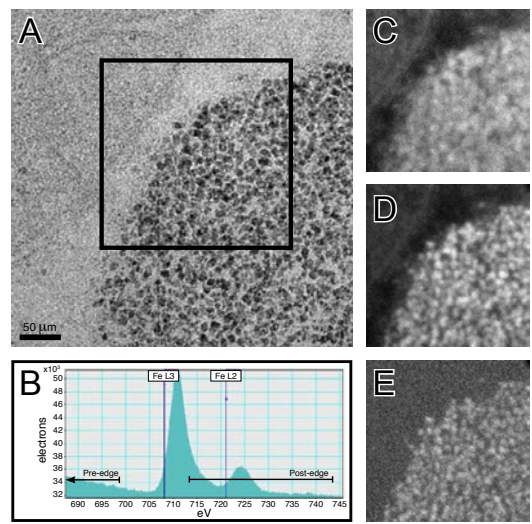

*Figure 5* Electron Energy Loss Spectroscopy (EELS) confirmed the dark spots accumulating in the liver and kidney to be γ-Fe$_2$O$_3$ NP. A: Transmission electron microscopy picture of the evaluated sample. B: EELS spectrum from the sample with indications of the position of the Fe-L2 and Fe-L3 and the position of the pre-edge and post-edge slit recordings. The edges in the EELS spectrum are consistent with Fe being the atom resulting in the electron energy loss. C: Image showing the positional origin in the sample of electrons in the pre-edge energy window. D: Image showing the positional origin in the sample of electrons in the post-edge energy window. E: Image showing the jump-ratio, indicating the positional origin of the Fe specific electrons in the sample. These results confirm the imaged particles to be the injected NP's

*Mean arterial blood pressure (MAP) measured by fluid-filed catheters*

The resting mean arterial pressure (MAP) measured prior to PAA coated γ-Fe$_2$O$_3$ NPs or saline IV injections, respectively, averaged 124.1±3.8 and 117.3±3.7 mmHg and the heart rate ($f_H$) 606±11 and 577±12 min$^{-1}$. NP injections caused a significant decrease in MAP (103.7±5.0 and 99.4±4.5 mmHg) 12-24h after administration (Fig. 6). This was associated with a tendency towards a decreased $f_H$, although this was not significant (483±64 and 487±58 min$^{-1}$). After 36h, both blood pressure and heart rate were restored.

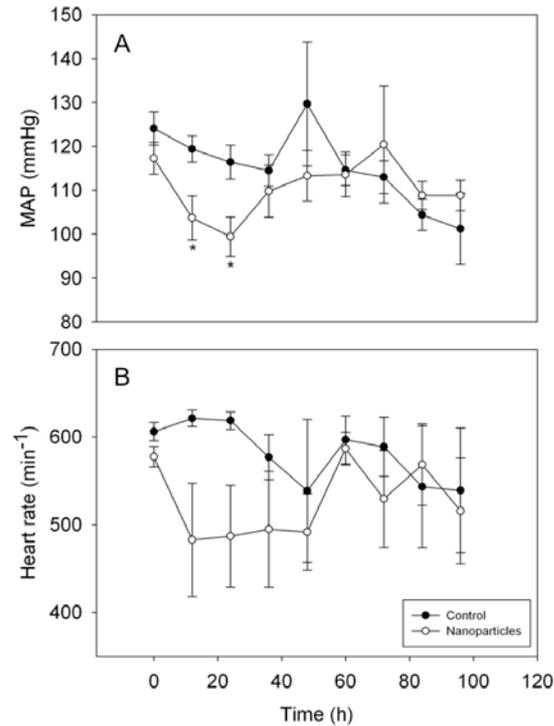

*Figure 6* Mean arterial blood pressure (MAP) and heart rate in healthy BLAB/c mice receiving PAA coated γ-Fe$_2$O$_3$ NP (10 mg kg$^{-1}$). The NPs were given at time 0. The filled circles represent control mice that received saline (n=9) where as the open circles represents mice that received NP (n=9). Asterisks (*) denote a significant difference (p < 0.05) compared to the control group. All data are shown as mean±SEM.

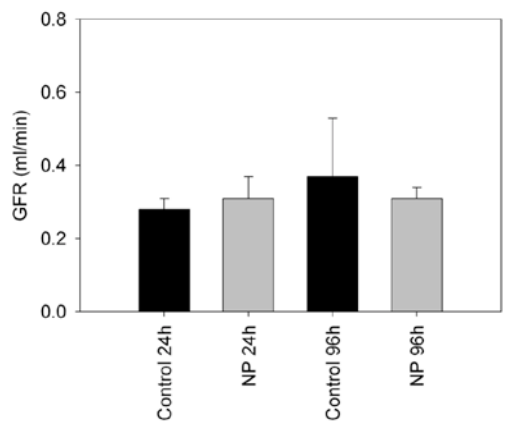

*Figure 7* Glomerular filtration rate (GFR) measured in two groups of mice receiving PAA coated γ-Fe$_2$O$_3$ NP (grey bars, n=9) and control mice (black bars, n=9) determined 24h and 96h after NP administration. Data are shown as mean±SEM.

*Glomerular filtration rate (GFR)*

TEM of the kidney revealed NP to accumulate in the mesangial cells and podocytes. Since these cells play pivotal roles in determining the glomerular filtration rate

(GFR), which is the classically used parameter to evaluate overall kidney function, we determined GFR in mice subjected to NP injections. The mice received intravenously injected PAA coated γ-$Fe_2O_3$ NPs following 24 and 96h. GFR was 11.0±0.8 and 13.0±0.4 ml $min^{-1}$ $g^{-1}$ following 24 and 96h of NP injections, and was not significantly different from the values obtained in mice receiving saline, serving as control group (14.0±0.2 and 14.0±0.6 ml $min^{-1}$ $g^{-1}$; Fig. 7).

*Telemetric blood pressure measurements*

The example of continuous measurements of mean arterial blood pressure (MAP), heart rate ($f_H$), and physical activity obtained 8 days after surgery (Fig. 8) reveal the typical murine circadian rhythms with elevated and more variable MAP, $f_H$ and activity during dark periods. To account for diurnal rhythms and activity-related changes, the data for MAP and $f_H$ were analyzed in different ways; i) the average model, and ii) the model proposed by Van Vliet et al. (2006) excluding the effect of light and activity.

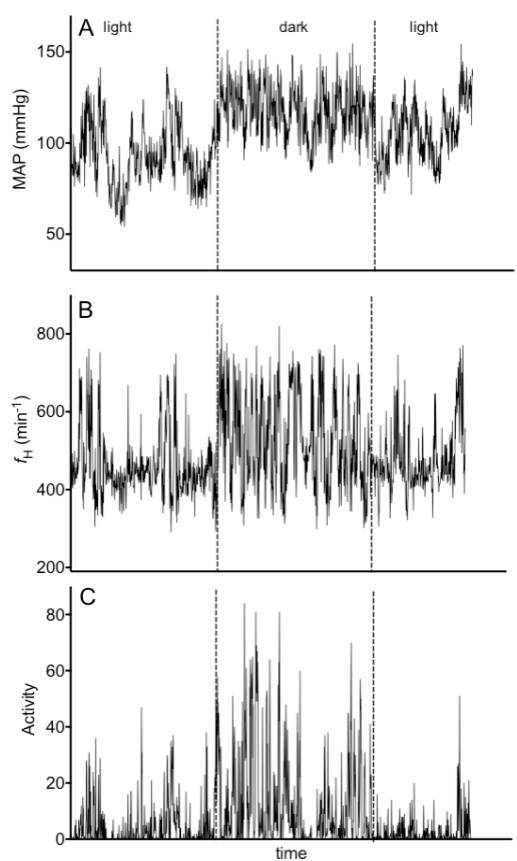

*Figure 8 Representative 36h data trace from one healthy BALB/c mouse 10 days after instrumentation. A: Mean arterial blood pressure (MAP), B: heart rate ($f_H$), C: locomotor activity. The dotted lines indicate the shift from light to dark-phases, respectively.*

Figure 9 presents the data analyzed with the first method when MAP, $f_H$ and activity were determined as the average of the three lowest values for each experimental group. Data are presented separately for the 12h light periods over the seven experimental days (left-handed panels) and the 12h dark periods (right-handed panels). In both the NP-treated group and the control group, MAP tended to decrease during the light periods (Fig. 9 A), although not statistically significant. MAP was not affected of time after treatment in either the control group receiving saline or in the mice, which received NP. Similar to MAP, $f_H$ was generally not affected by NPs (Fig. 9 C, D). In both groups, both during light and dark phases, there was a tendency, although insignificantly, of decreased $f_H$ over the 7 days. There was no effect of NP injections on locomotor activity (Fig. 9 E, F).

In the second analysis, we used the method of Van Vliet et al (2006) to calculate average MAP as:

$$MAP_{24h} = MAP_{inactive} + T_{active} (MAP_{active} - MAP_{inactive}),$$

to exclude the effect of circadian rhythms and activity on MAP (e.g. Figs 8), where $MAP_{24h}$ represents mean MAP over the entire 24 h period, $MAP_{inactive}$ and $MAP_{active}$ the MAP during inactivity (no activity signal) and activity, respectively. $T_{active}$ represents the proportion of time spent in activity. By emphasizing the impact of activity, the calculated MAP depends on (i) the extent of the change in $T_{active}$ and (ii) the magnitude of the effect of activity on MAP ($MAP_{active} - MAP_{inactive}$) (Van Vliet et al. 2006). Figure 10 shows data from the present study analyzed based on this method. When compensating for activity level, there was a significant reduction in MAP on the second and seventh day after intravenous injection of $\gamma$-$Fe_2O_3$ NPs, while here was no observed effect of injecting saline.

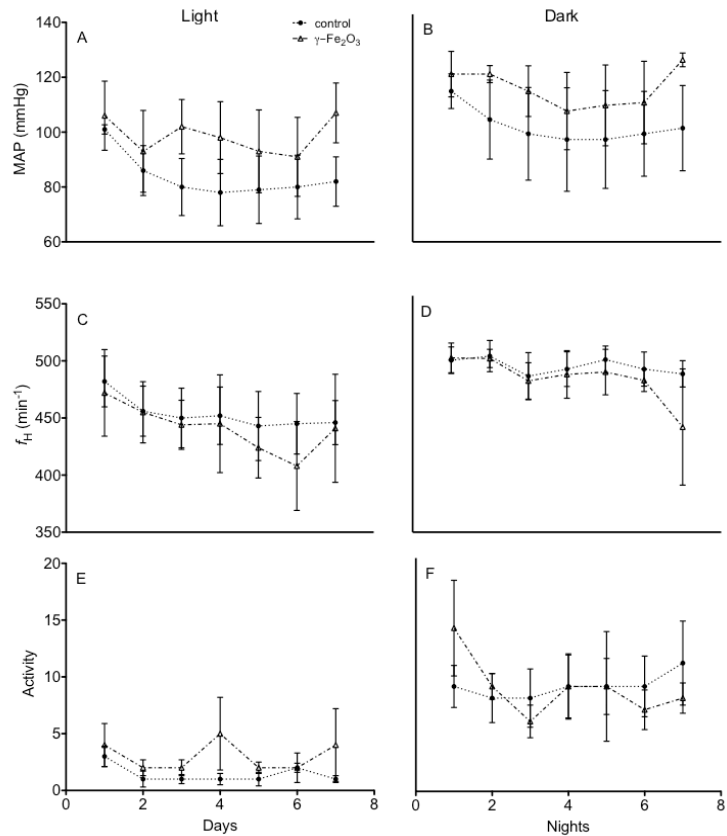

*Figure 9* Mean arterial blood pressure (MAP), heart rate ($f_H$) and activity of the two experimental groups of healthy BALB/c mice that received PAA coated $\gamma$-$Fe_2O_3$ NP ($\Delta$, n=6) and saline (•, n=6). Data are presented as mean±SEM.

*Contractile properties of mesenteric vessels in vitro*

Since no renofunctional effects were seen in mice receiving NP injections, the drop in MAP observed prompted us to investigate the vascular contractility. All mesenteric arteries exhibited a concentration-dependent contraction in response to noradrenaline (NA). However, the maximal response to NA was significantly lower in arteries from mice injected with iron oxide NP and *in vitro* treated vessels at the lowest dose. Furthermore, the maximal response to NA tended to decrease at the highest dose of $\gamma$-$Fe_2O_3$ (Fig. 11 A). This was similar to the mesenteric vessels from mice that received $\gamma$-$Fe_2O_3$ NPs 24 and 72 h prior to the myograph measurements, whereas vessels from mice exposed to iron oxide for 7 days did not differ from control mice (Fig. 11 B, C, D). In addition, the maximal force developed in the $\gamma$-$Fe_2O_3$ NPs *in vitro* protocol significantly failed to develop as high as the value obtained in the control groups (Fig. 11 A). There was only an effect of γ-$Fe_2O_3$ NPs *in vivo* exposure on pressure development after 24h of exposure (Fig. 12 B, C, D).

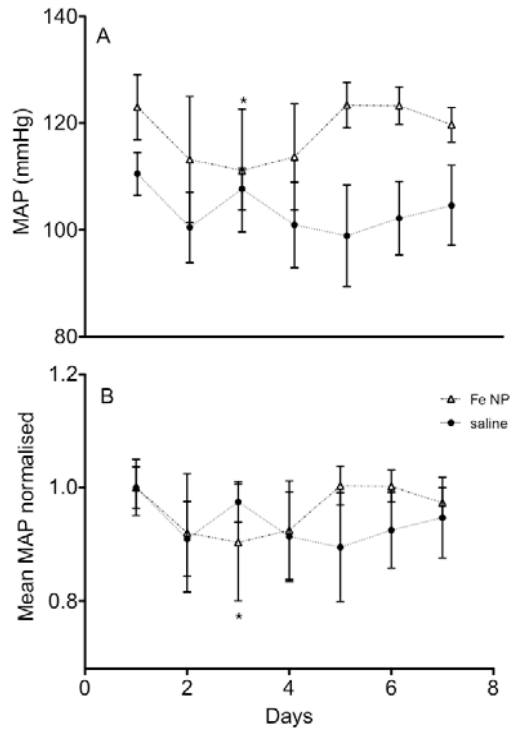

*Figure 10* Mean arterial blood pressure (MAP) in the two experimental groups of mice that received PAA coated γ-$Fe_2O_3$ NP (Δ, n=6) and saline (•, n=6). A; calculated from Van Vliet et al. (2006), and B; when data was normalised to the control value in each group. Data are presented as mean±SE. Asterisks (*) denote a significant compared to the control group, dobble dagger (‡) indicates significant difference ($p < 0.05$) between the two treatment groups.

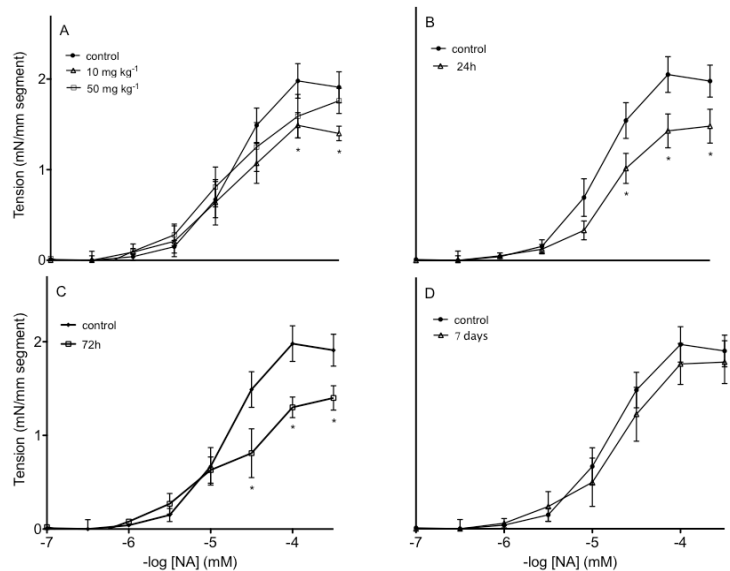

*Figure 11* Cumulative dose-response curves of tension of small mesenteric arteries from healthy BALB/c mice after receiving PAA coated γ-$Fe_2O_3$ NP (Δ) and saline (•) to increasing concentrations of noradrenaline (NA). A; in vitro study with γ-$Fe_2O_3$ NP (n=18), B; intravenous injected PAA coated γ-$Fe_2O_3$ NP (Δ) and saline (•) for 24h (n=6+6), C; 72h (n=6+6) and D; 7 days (n=6+6). Asterisks (*) denote a significant difference ($p < 0.05$) compared to the control group. Data are presented as mean±SEM.

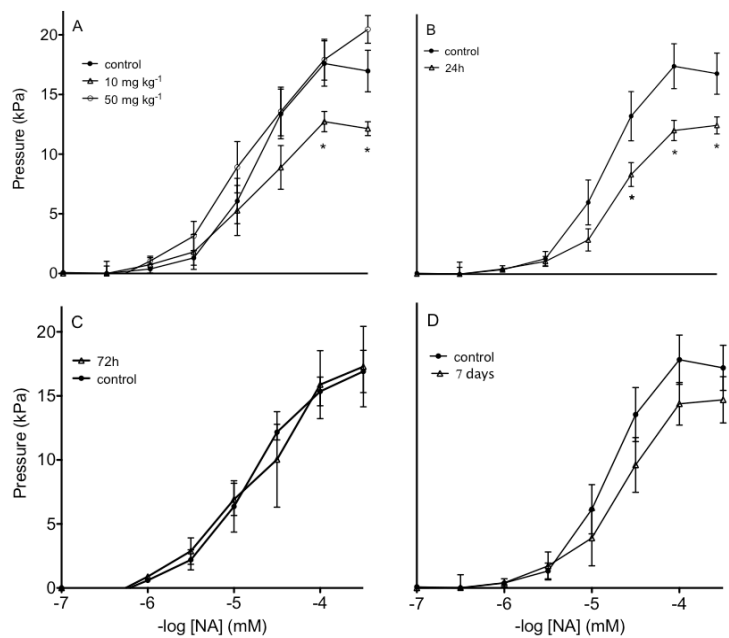

*Figure 12* Cumulative dose-response curves of pressure development of small mesenteric arteries from healthy BALB/c mice after receiving PAA coated γ-Fe$_2$O$_3$ NP (Δ) and saline (•) to increasing concentrations of noradrenaline (NA). A; in vitro study with γ-Fe$_2$O$_3$ NP (n=18), B; intravenous injected PAA coated γ-Fe$_2$O$_3$ NP (Δ) and saline (•) for 24h (n=6+6), D; for 72h (n=6+6) and C; for 7 days (n=6+6). Asterisks (*) denote a significant difference ($p < 0.05$) compared to the control group. Data are presented as mean±SEM.

## Discussion

In agreement with previous studies, the PAA coated γ-Fe$_2$O$_3$ NPs (Fig. 1) accumulated rapidly in liver upon intravenous injection, followed by a slower, but pronounced accumulation in the kidneys (Fig. 2 B, F). This temporal pattern of iron oxide NP seems to be independent of size, coatings, and route of administration (Briley-Saebo et al., 2004; Jain et al., 2007; Zhu et al., 2008; Thakor et al., 2011). In our study, the NPs in the liver were not incorporated in the hepatocytes, but appeared very clearly in both endothelial and Kupffer cells in the liver sinusoids. Given the generally very high permeability of the endothelial cells in liver sinusoids, it is surprising not to see NPs in hepatocytes. However, this indicates that the Kupffer cells and endothelial cells of liver sinusoids perform a filtering function, which may protect the hepatocytes from particles. Previous studies have also shown that iron oxide particles are phagocytized or endocytized by the immune cells of the reticular endothelial system (RES) in liver, spleen, lymph and bone marrow (Wisner et al., 1995; Chouly et al., 1996; Rety et al., 2000; Allkemper et al., 2002). However, despite the rapid accumulation of NPs in macrophages and clearance by the RES system, the particles may cause temporary inflammation and apoptosis in the liver (Thakor et al., 2011). Furthermore, the TEM images showed that the PAA coated γ-Fe$_2$O$_3$ NPs appeared in vesicular structures of mesangial cells, podocytes and proximal tubular cells within the kidney (Fig. 4). When the same PAA coated γ-Fe$_2$O$_3$ NPs are incubated in lymphoblast cells *in vitro*, the NPs were absorbed onto the membrane and in the cytoplasm in the membrane bound compartments (Safi et al., 2011). In our study, NPs were not found in the cytosol or in the nucleus. The slow accumulation in the kidney in spite of efficient build up in liver endothelial and Kuppfer cells may reflect an iron NP overload of the hepatic RES.

Because of the buildup of the PAA coated γ-Fe$_2$O$_3$ NPs in podocytes, we investigated whether the metal NP may affect kidney function. It has been discussed whether the kidneys excrete intravenously injected NPs, by either glomerular filtration and/or tubular excretion (Longmire et al., 2008). Neither mesangial cells nor podocytes has direct access to NP's in blood plasma, so the presence of NP's in these cells show that the NP's pass the endothelial lining of the glomerular capillaries. These endothelial cells are of the fenestrated type, and may thus allow NP's to pass to the interstitial space, and be accessible to mesangial cells. The presence of NPs in podocytes moreover indicates that NPs can pass the glomerular basement membrane, and thus are filtered into the urine. This is further supported by the finding of NPs in endocytic vesicles/lysosomes in the cytoplasm of proximal tubular cells. The proximal tubule connects Bowman's capsule to the loop of Henle and is of major importance for reclamation of filtered salt, water, bicarbonate, glucose and proteins as well as for secretion of organic acids, such as creatinine and other bases. In spite of the NP accumulation in this part of the kidney tubules, there was no disturbance of plasma [HCO$_3^-$] or arterial acid-base balance upon intravenous injections of NP. Thus, it seems that the accumulated NPs do not affect renal acid-base regulation. The control values for plasma [HCO$_3^-$] as well as arterial pH and PCO$_2$ were similar to previous measurements on chronically instrumented mice (Wall et al., 2004; Chambrey, 2005; Vallet, 2006; Hafner et al., 2008; Iversen et al., 2012). Importantly, we also demonstrated that the accumulation of iron oxide NPs in the kidneys did not affect GFR (Fig. 7). This indicates that despite NP accumulation in podocytes and glomerular mesangial-cells, the essential functions of these cells in regulation of GFR are still intact.

*Cardiovascular effects*

Although exposure to ultrafine particles (UFP; less than 1 µm) with similar physiochemical properties as NPs is known to cause adverse cardiovascular effects, (Hamor et al., 2003; Pope et al., 2002, 2004; Stern and McNeil, 2008) only a few previous studies have addressed the question of how engineered NPs affect the cardiovascular system (Apte and Hede, 2007; Pope et al., 2004). In our study, the transient decrease in MAP at 12-24h after NP administration (Fig. 6, 10) may be associated with endothelial dysfunction due to oxidative stress as hypothesized by Mo et al. (2009) and further may be explained by the uptake of NP in endothelial cells demonstrated in the present study (Fig. 3 A, B). Telemetry allows for MAP to be measured accurately under relatively stress-free conditions in freely moving, conscious animals, thus alleviating the effects of stress from recent anesthesia and surgery (Butz and Davisson, 2001). The circadian rhythms in MAP, $f_H$ and activity level were similar to previous studies using telemetry in mice Van Vliet et al., 2006) and in accordance with these studies, we also found pronounced individual variation. The rise in MAP during activity appears to be sympathetically mediated (Swoap et al., 2004) and locomotor activity is considered to be the major determinant of the variation in MAP amongst individuals. This may explain the significantly elevated MAP during the dark periods in mice receiving γ-Fe$_2$O$_3$ NPs.

We have demonstrated that acute decrease in MAP followed γ-Fe$_2$O$_3$ NPs injections. The effects of the injected NP on MAP are consistent with previous reports on inhalation of UFPs, as well as inhaled and injected NPs (Pope et al., 2004). Intravenous injections of gold nanoparticles in rats lead to a decreased QRS complex and increased time period between two repolarisation phases (Apte and Hede, 2007). In comparison, Albini et al. (2010) showed that single-wall carbon nanotubes were easily taken up by endothelial cells *in vitro* and were associated with an enhanced acidic vesicle compartment within the cells. In the myograph study addressing the endothelial function after NP administration, small mesenteric arteries from healthy mice incubated with PAA coated γ-Fe$_2$O$_3$ NP *in vitro* developed a reduced maximal forced in response to increasing concentrations of noradrenaline. Similarly, small mesenteric arteries from mice, which had received the iron oxide NPs for 24h *in vivo* had a reduced maximal force of contraction, whereas there was no effect on the endothelial function after 7 days of NP exposure. In comparison, Apte and Hede (2007) showed that intravenous injections of gold NP (1 μg kg$^{-1}$) caused mild bradycardia in rats with $f_H$ decreasing from 413 to 387 min$^{-1}$ at a dose one to tenth of the suggested dose for normal clinical settings. However, other theories have been presented to explain these cardiovascular effects of UFP's. Instillation of single-walled carbon nanotubes (SWCNTs; 1 μg g$^{-1}$) affected autonomic control of blood pressure following four weeks of telemetric measurements in rats. This finding was ascribed to induction of a peculiar inflammatory pulmonary reaction (Legramante et al., 2009). In a similar study, Thakor et al. (2011) reported no change in ECG, MAP or heart rate two weeks after intravenous or intraperitoneal administration of Raman-silica-gold-NP (R-Si-Au-NP; 200 μl of stock solution with 9.6 * 10$^{10}$ NP). The cardiovascular parameters, however, were only reported on a weekly basis in the study of Thakor et al, (2011). In our study, the transient reduction in MAP was only observed at 12-24 h, and thus would not have been identified in the study by Thakor et al. (2011). We find it likely that the NPs are rapidly removed from the systemic circulation by the RES thereby limiting the potential toxic effects. The rapid uptake of NPs in the liver RES cells shown in our study may result from the transient hypotension shortly after administration and be the mechanism, which allows a normalisation of MAP. In another study, hypertensive rats exposed to instilled UFPs increased $f_H$ and MAP, whereas the normal rats did not respond (Gordon et al., 1998). The authors therefore hypothesized that inhaled PM causes systemic effects by evoking a stress that may compensated in healthy individuals, but that severely disrupts homeostatic processes in individuals with cardiovascular or pulmonary diseases (Gordon et al., 1998). This underpins the importance of future studies on the cardiovascular effects of NPs during disease states, because most patients to receive NP are likely to suffer from various diseases, including hypertension.

*The cause of the NP effect –endothelial stress?*

The effects of NPs on the cardiovascular system have been proposed to relate to the formation of oxidative stress that impairs normal endothelial function (Behrendt et al., 2002). We have shown in the present study that the γ-Fe$_2$O$_3$ NPs are taken up by endothelial cells, particularly in liver and kidney, and that the small mesenteric arteries exposed short-term to these γ-Fe$_2$O$_3$ NPs failed to obtain the maximal force development when stimulated with NA. This may support the proposed hypothesis. The arteries used were small enough to be regarded as resistance arteries and are therefore considered to be of importance for the regulation of flow distribution (Mulvany and Halpern, 1977). The main observation is that the major endothelial-dependent relaxing pathways seem to be intact, though the contractile response to NA is decreased. This may be due to the development of oxidative stress. Mo et al. (2009) showed that UFPs, at a non-toxic dose, induced reactive oxygen species (ROS) generation in mouse pulmonary micro vascular endothelial cells. In the present study, both the short-term (24h and 72h) exposure *in vivo* to γ-Fe$_2$O$_3$ NPs and the 0.5h *in vitro* exposure affected the endothelial function, so generation of ROS seems unlikely during our experimental conditions due to the short time of exposure. Therefore, the effects on endothelial function observed here may be explained by an uptake of the NP in the endothelial cells. Thus, endothelial cells, which line the inner surface of blood vessels, are in direct contact with these particles, emphasizing the relevance of particle-endothelial interactions. If a pro-inflammatory stimulation of endothelial cells by nano-scaled particles occurs *in vivo*, a chronic inflammation (such as granulomatosis) could be a possible consequence. Since endothelial proliferation is a prerequisite for blood vessel formation (e.g. during wound healing), an impairment of endothelial proliferation could indicate a reduced capacity for blood vessel formation (angiogenesis*) in vivo* (Peters et al., 2004).

We demonstrate that intravenous injections of γ-Fe$_2$O$_3$ NPs NPs cause an acute reduction in MAP, compensated for by cardiovascular adjustments or a termination of the

NP effect.

**Acknowledgements.** This study was supported by EU-FP7 (FP7-NMP-2007-SMALL-1) and The Danish National Science Research Council